\documentclass[aps,prd,nofootinbib,superscriptaddress,twocolumn]{revtex4-1}

\usepackage[bookmarks=true]{hyperref}  
\usepackage{amsmath,amsfonts,amsmath,ulem}
\usepackage{graphicx,subfigure,epsfig,color}

\newcommand{\ba}{\begin{eqnarray}}
\newcommand{\ea}{\end{eqnarray}} 
\newcommand{\nn}{\nonumber}

\def \be {\begin{equation}}
\def \ee {\end{equation}}
\def \barr {\begin{array}}
\def \earr {\end{array}}
\def \bea {\begin{eqnarray}}
\def \eea {\end{eqnarray}}

\def \ble {\begin{widetext}\begin{equation}}
\def \ele {\end{equation}\end{widetext}}
\def \blea {\begin{widetext}\begin{eqnarray}}
\def \elea {\end{eqnarray}\end{widetext}}

\def \nn {\nonumber}

\newcommand{\eq}[1]{(\ref{#1})}

\def \and {{\textrm{and}}}

\begin{document}

\title{Hawking Flux from a Black Hole with Nonlinear Supertranslation Hair}

\author{Feng-Li Lin}
\email{linfl(at)gapps.ntnu.edu.tw} 
\affiliation{Department of Physics, National Taiwan Normal University, Taipei 11677, Taiwan}

\author{Shingo Takeuchi}
\email{shingo.portable(at)gmail.com}
\affiliation{Phenikaa Institute for Advanced Study and Faculty of Basic Science, Phenikaa University, Hanoi 100000, Vietnam}

\begin{abstract}
We study the Hawking flux from a black hole with soft hair by the anomaly cancellation method proposed by Robinson and Wilczek. Unlike the earlier studies considering the black hole with linear supertranslation hair, our study takes into account the supertranslation hair to the quadratic order, which then yields the angular dependent horizon.
As a result, highly nontrivial kinetic-mixings appear among the spherical Kaluza-Klein modes of the (1+1)d near-horizon reduced theory, which obscures the traditional derivation of the Hawking flux. However, after a series of field re-definitions, we can disentangle the mode-mixings into canonical normal modes, but the reduced metrics for these normal modes are mode-dependent. Despite of this, the resultant Hawking flux turns out to be mode-independent and remains the same as the Schwarzschild's one. Thus, one cannot tell the black holes with nonlinear supertranslation hairs from the Schwarzschild's one by examining the Hawking flux, so that the nonlinear soft hairs can be thought as the microstates.

\end{abstract}

\maketitle


\section{Introduction}
Black hole information paradox \cite{Mathur:2009hf} invoked by the discovery of Hawking radiation \cite{Hawking:1974rv,Hawking:1974sw} 
indicates our insufficient understanding of quantum gravity and the nature of microstates accounting for the Bekenstein-Hawking entropy \cite{Bekenstein:1973ur}. 
The microstates by definition should share the common macroscopic observables, i.e., the so-called hard hairs, which in black hole physics is the mass, charge and angular momentum, and equivalently 
the Hawking temperature and Hawking flux in the thermodynamic sense. Therefore, the additional degrees of freedom accounting for the black hole microstates can only be the so-called soft hairs, which will 
not change the hard hairs but may change the near-horizon properties such as it shape and local geometry.   
Accordingly, the soft hairs proposed in \cite{Hawking:2016msc,Strominger:2014pwa,Strominger:2017zoo} can be microstates if they also keep the thermal properties of black hole intact. The soft hairs are the conserved charges of the infinite-dimensional symmetries, i.e., the BMS symmetries \cite{Bondi:1962px,Sachs:1962wk,Sachs:1962zza} of the asymptotic flat spacetime, and can be observed through the gravitational memory effect \cite{Braginsky:1986ia,Braginsky&Thorne,Blanchet:1992br}. Moreover,  the soft hairs being treated as microstates are also proposed in \cite{Hawking:2016msc} as a resolution to information paradox.  The key idea of \cite{Hawking:2016msc} is to view the formation and evaporation of a black hole as the scattering process, the infinite relations imposed by BMS symmetries are argued in \cite{Hawking:2016msc} to be used to map the soft hairs of the collapsing matters into the outgoing Hawking flux, and thus help to resolve the information paradox. However, the further analysis \cite{Mirbabayi:2016axw,Bousso:2017dny} showed that the soft modes are decoupled from the scattering of the hard modes after appropriately dressing the in and out states. This will preserve the Hawking flux and thus the microstate nature of the {\it linear} soft hairs, as shown in \cite{Javadinazhed:2018mle}. 
 
As far as we know, only two examples of exact soft-hairy black hole solutions 
have been constructed. One is considered by Hawking, Perry and Strominger (HPS) \cite{Hawking:2016sgy}  
by throwing the soft-hair shockwave into the Schwarzschild black hole, and the other is 
explicitly constructed by Compere and Long (CL) \cite{Compere:2016hzt} through 
nontrivial coordinate transformations. Their main difference is that the HPS black hole 
carries linear supertranslation hair but the CL's one  carries nonlinear one. We will show 
in this study, the nonlinear effect will deform the horizon into non-spherical shape and 
this non-spherical horizon will obscure the derivation of the Hawking  flux. 

As shown in the next section, the surface gravity of
the supertranslated black hole remains the same with the
Schwarzschild's one. This implies the same Hawking temperature and possibly the same Hawking flux. However,
the non-spherical horizon makes the task of evaluating
Hawking flux nontrivial even if the final result remains
the same as Schwarzschild case. Therefore, we in this paper will try to resolve the issue of evaluating the Hawking
flux for the non-spherical horizon, and clarify the role of
nonlinear soft hairs as microstates.


 There are various ways to derive the Hawking radiation from conventional black holes by considering the quantum field theory (QFT) in curved space. Besides Hawking's original derivation \cite{Hawking:1974sw}, Wilczek and his collaborators have developed two alternative methods of deriving the Hawking radiation. First, a tunneling method proposed by Parikh and Wilczek \cite{Parikh:1999mf} is to calculate the WKB amplitude of classical forbidden path to evaluate the radiation spectrum. Later, motivated by the earlier trace-anomaly proposal \cite{Christensen:1977jc} for 2d gravity,  Wilczek and his collaborators \cite{Robinson:2005pd,Iso:2006wa,Iso:2006ut} utilize the chiral nature of the QFT in the reduced (1+1)d  near horizon region due to the absence of ingoing modes, which results in the (consistent) gravitational anomaly \cite{AlvarezGaume:1983ig,Bertlmann:2000da}
\be
\nabla_\mu T^\mu{}_\nu =  \frac{1}{\sqrt{-g}}\partial_{\mu} N^{\mu}_{\nu}
\ee 
where for the Schwarzschild-like metric $ds^2=-f(r) dt^2+ dr^2/f(r)$, the non-vanishing components of $N^{\mu}_{\nu}$ is
\be\label{Nmunu}
N^r_t={1\over 192}(f'^2+f''f), \quad N^t_r={-1\over 192 f^2}(f'^2-f''f).
\ee
Thus, an extra flux, i.e., the Hawking flux, is induced to cancel the gravitational anomaly so that the underlying covariance is preserved.

The Hawking flux for the HPS black hole and its Vaidya type generalization has been investigated in \cite{Chu:2018tzu} by adopting the tunneling method of \cite{Parikh:1999mf}, 
and in \cite{Compere:2019rof} (see also \cite{Javadinazhed:2018mle}) by using Hawking's original method. Both conclude that the Hawking flux from HPS's remains the same as 
the Schwarzschild's one. In this letter we will consider this issue for the CL's one by anomaly cancellation method of \cite{Robinson:2005pd}.  As mentioned, the non-spherical 
horizon deformed by the nonlinear soft hair poses a challenge to examine the thermal nature of its Hawking flux. Indeed, the complication arises when performing the dimensional 
reduction since the non-spherical horizon shape induces nontrivial mode-mixings in the (1+1)d reduced theory, which prevent identifying the normal modes and obtaining their 
associated Hawking flux. Despite of this, we will show that the mode-mixings can be removed by nontrivial field-redefinitions so that the resultant Hawking flux is still the same as 
the Schwarzschild's one.      
  
  The remaining of the paper is organized as follows. In the next section, we expand the metric for the static soft-hairy black hole \cite{Compere:2019rof} up to the second order of the soft hair function, then transform it into the Schwarzschild coordinate for a chosen soft hair function and check its surface gravity to be the same as the one without soft har. In section \ref{sec 3} we perform the dimensional reduction to obtain the near-horizon reduced theory and spell out the structure of mode-mixings. In section \ref{sec 4} we remove the infinite mode-mixings by a series of nontrivial field re-definitions, and then arrive the diagonal reduced theory, and based on which,  in section \ref{sec 5} for each ``normal mode"  we obtain the Hawking temperature from the reduced metric, and the Hawking flux by the method of anomaly cancellation. We find the results are the same as the ones for Schwarzschild black hole. In section \ref{sec 6} we conclude our paper and discuss why our results should hold for general quadratic soft hair.  Besides, in Appendix \ref{app A} we collect the details of the metric of supertranslated black hole considered in this paper, and in Appendix \ref{app B} we collect the full details of the near-horizon reduced theory, which is the starting point up to some field re-definitions for the method of anomaly cancellation to derive the Hawking flux\footnote{Detailed calculation process in this paper is given in \cite{Takeuchi:2021ibg}.}.

\section{Black holes with supertranslation soft-hairs} \label{sec 2}
Supertranslations are the direction-dependent time-like Killing transformations on the null-like infinities of the asymptotically flat spacetimes, the metric of which in the Bondi coordinates specifically for the past null infinity ${\cal I}^-$ takes the following form
 \bea
 ds^2&=&-dv^2 + 2dv dr + r^2 \gamma_{AB} dz^A dz^B
\nn \\
 &+& {2M\over r}dv^2+rC_{AB}dz^A dz^B-D^BC_{AB} dv dz^A +\cdots \qquad
 \eea
where $\gamma_{AB}$ is the metric of unit 2-sphere, $D_A$ is the covariant derivative with respect to $\gamma_{AB}$. The Bondi mass aspect $M$ and the traceless tensor $C_{AB}$ depend on $(v,z^A)$, and characterize the high degeneracy of classical vacua of general relativity. The associated Killing field preserving the above asymptotic metric in the Bondi gauge is as follows:
\be\label{Killing}
\xi_C=C \partial_v-{1\over r}D^AC \partial_A+{1\over 2} D^2C\partial_r
\ee
where $C=C(z^A)$. We can also obtain the Bondi metric for the the future null infinity ${\cal I}^+$ in a similar way in terms of the outgoing Eddington-Finkelstein coordinate $(u,r,z^A)$. 

   What we will be interested is the CL type supertranslated black hole which is the vacuum solution of Einstein equation and carries the supertranslation charges of the above asymptotic BMS symmetries. Its metric in the isotropic coordinate 
takes the following form \cite{Compere:2019rof}:
\bea 
ds^2&&=-\frac{\left( 1 - \frac{M}{2\rho_s} \right)^2}{\left( 1 + \frac{M}{2\rho_s} \right)^2}dt^2 +\left( 1 + \frac{M}{2\rho_s} \right)^4 \times \nn \\
\label{fullBHs}
&&  \left( d\rho +[((\rho-E)^2+U)\gamma_{AB}+(\rho-E)C_{AB}]dz^A dz^B \right) \qquad 
\eea
where the auxiliary radial coordinate $\rho_s$ is related to the radial coordinate $\rho$ by $\rho_s=\sqrt{(\rho-C-C_{(0,0)})^2+D_ACD^AC}$, here $C_{(0,0)}$ denotes the constant mode of $C=C(z^A)$ in spherical harmonic expansion. Besides, 
\bea
&& C_{AB}:=-(2D_AD_B-\gamma_{AB} D^2)C,  \nn \\
&& U:={1\over 8} C_{AB}C^{AB}, \qquad E:={1\over 2} D^2C+C-C_{(0,0)}.
\eea
This metric reduces to Schwarzschild black hole of mass $M$ in the isotropic coordinate when setting $C=0$. 
   
  Due to the nontrivial relation between $\rho$ and and $\rho_s$, the metric \eq{fullBHs} is to all orders of $C$ in $C$-expansion, thus the supertranslation hair here is nonlinear in contrast to the HPS's one. Up to the first order of $C$, $\rho=\rho_s+C+C_{(0,0)}+{\cal O}(C^2)$ so that $d\rho=d\rho_s+\partial_A C dz^A$, then we can transform the metric \eq{fullBHs} up to the first order of $C$ to the Schwarzschild coordinate $(t,r,z^A)$ by the relation $r=\rho_s(1+{M\over 2\rho_s})^2$. The resultant metric has the same $tt$- and $rr$-components as the Schwarzschild metric so that the position of the horizon and surface gravity are the same as the ones of Schwarzschild black hole. One expects the Hawking flux is thermal as usual. This result is the same as the one obtained in \cite{Chu:2018tzu} for the HPS black which also carries just the linear supertranslation hair. 
 
  We now expand the metric to the second order of $C$, and examine the same issue.  For concreteness and simplicity, we consider the following soft hair\footnote{We chose this mode because we expect it will dominate when forming a soft-hairy black hole from the coalescence of binary black holes due to the evidences from the empirical studies of the emitted  gravitational wave in the merger and ringdown phases, see e.g. \cite{Berti:2005ys}}:
\be\label{Cfunction}
 C = \varepsilon M Y_2^0 (\theta,\phi)= \varepsilon M \sqrt{\frac{5}{16\pi}}\left(3\cos^2\theta-1\right) 
 \ee
so that $C_{(0,0)} = 0$.  Here, $\varepsilon$ is a small parameter to keep track of the order in the $C$ expansion of the metric. The factor of $M$ in \eq{Cfunction} is just our convention. Moreover, to avoid lengthy formulas, we will just write down the metric in the near-horizon region relevant for the anomaly cancellation method of deriving Hawking flux. The full metric is given in the Appendix \ref{app A}. The resultant near-horizon metric in the Schwarzschild (not the isotropic) coordinate up to ${\cal O}(r-r_h)$ and ${\cal O}(\varepsilon^2)$ is
\be\label{metricB}
ds^2=g_{tt}dt^2+g_{rr}dr^2+2 g_{r\theta}dr d\theta+ g_{\theta\theta} d\theta^2 + g_{\phi\phi} d\phi^2 \nn
\ee
with
\bea \label{gtt-nh}
g_{tt}&=&-g_{rr}^{-1}=-{r-r_h \over 2M} (1-{45 \varepsilon^2 \sin^2 2\theta \over 8\pi}),
\\
g_{\theta\theta}&=& r^2 \Big(1+ 6\sqrt{5\over \pi} \varepsilon \cos 2\theta +{5\varepsilon^2 \over 16\pi} (75+69 \cos 4\theta)\Big),
\\
g_{r\theta}&=&-3\varepsilon \sqrt{10 M^3 \over \pi (r-r_h)} \sin 2\theta + {15 \varepsilon^2 M^2 \sin 4\theta \over 2\pi (r-r_h)},
\\ \label{gphiphi-nh}
g_{\phi\phi}&=&r^2 \sin ^2 \theta \Big(1+ 3 \varepsilon \cos^2 \theta \Big\{ 2\sqrt{5\over \pi} + {5\over 4 \pi} \varepsilon \nn \\
&&- {35 \over 8\pi}  (r-r_h) \varepsilon + {55\over 4\pi} \varepsilon \cos 2\theta \Big\} \Big)
\eea
and 
\be
r_h= 2M(1-{15 \varepsilon^2 \sin^2 2\theta \over 16\pi}).
\ee

 Note that $g_{tt}=0$ at $r=r_h$ so that the horizon position is shifted from $2M$ to $r_h$, and the more important it is now angular dependent. To make sure $r=r_h$ is indeed a null hypersurface $\cal N$, we evaluate the norm of the normal vector $\ell$ to a family of surfaces $S=r-r_h$ \cite{Townsend:1997ku}, i.e.,
\be
\ell^2\propto g^{\mu\nu} \partial_{\mu}S\partial_{\nu} S
\ee
 and find that $\ell^2$ indeed vanishes at $r=r_h+{\cal O}(\varepsilon^3)$. We then evaluate the surface gravity $\kappa$ for the time-like Killing vector $\xi=\partial_t$ (for static solutions) \cite{Townsend:1997ku} and obtain
 \be\label{kappaS}
 \kappa=\sqrt{-{1\over 2} (D^{\mu}\xi^{\nu})(D_{\mu}\xi_{\nu})\vert_{\cal N}}={1\over 4M}+{\cal O}(\varepsilon^3).
 \ee
 We see that the surface gravity equals to the one of Schwarzschild black hole even though the horizon is no longer spherical symmetric. Below we will calculate the associated Hawking flux to examine if the quadratic supertranslation hairs can be the microstates.

\section{Near-horizon reduced theory} \label{sec 3}
We will calculate the Hawking flux \`a la the anomaly cancellation method \cite{Robinson:2005pd}, by first finding the reduced action of a free scalar theory in the near-horizon region of the background \eq{metricB}
\bea
S&=&\int d^4x \sqrt{-g} g^{\mu\nu} \partial_{\mu}\phi\partial_{\nu}\phi^*,   \\
&=&\int d^4x \sqrt{-g} [ g^{tt} |\partial_t\phi|^2 +g^{rr}|(\partial_r+A_{\theta}\partial_{\theta})\phi|^2 \nn\\
&& \qquad \qquad -{(g^{r \theta})^2\over g^{rr} }|\partial_{\theta}\phi|^2]   
\eea
where the last term involving ${(g^{r \theta})^2\over g^{rr} }$ is sub-leading in the near horizon limit and can be neglected, and the gauge field $A_{\theta}\equiv {g^{r \theta}\over g^{rr} }$ is the $\theta$-component not the $t$-component so that it is irrelevant when considering the Hawking flux, thus we will omit it in what follows. 

To proceed further, we decompose the scalar field in the spherical harmonic expansion, i.e., $\phi=\sum_{\ell,m}\phi_{l m} Y_l^m$. Therefore, the relevant near-horizon action up to ${\cal O}(\varepsilon^2)$ is explicitly given by
\be
S_{n.h.}=\sum_{k,n,l,m} \int dt dr \int d\Omega\; \Lambda \; Y_k^{n*} Y_l^m (\bar{g}^{tt}|\partial_t \phi|^2 +\bar{g}^{rr}|\partial_r \phi|^2)  
\ee
where 
\bea
\bar{g}^{tt}&=&-(\bar{g}^{rr})^{-1}=-{2M\over r-2M}(1-{15 \varepsilon^2 M \sin^2 2\theta \over 8\pi (r-2M)}) \\
\Lambda&=&4M^2 \sin \theta \Big(1+12 \varepsilon Y_2^0+\varepsilon^2 ({120\over 7\sqrt{\pi}} Y_4^0+{54\over 7}\sqrt{5\over \pi} Y_2^0)\Big). \nn
\eea

After performing the spherical integral, the near-horizon action involves the mode-mixings and takes the following form:
\bea\label{mixingS}
S_{n.h.}&=&\sum_{k,n,l,m}\int d^2x\; 4M^2 \Lambda_{kn,lm} ((g_{\rm eff})^{tt}_{kn,lm} \partial_t\phi^*_{kn}\partial_t\phi_{lm} \nn\\
&& +(t\rightarrow r)) 
\eea
where 
\bea
\Lambda_{kn,lm}&=&\int d\Omega \; \Lambda \; Y_k^{n*} Y_l^m \nn \\
&\equiv& \Lambda_{lm}^{(0)} \delta_{kl}\delta_{mn} +  \Lambda_{lm}^{(2)}  \delta_{k-2,l}\delta_{mn} +   \Lambda_{lm}^{(4)}  \delta_{k-4,l}\delta_{mn} \qquad
\eea
with $\Lambda^{(K)}_{kn,lm}$ is a function of $k,n,l,m$ and ${\cal O}(\varepsilon^{K/2})$, and its explicit form is given in Appendix \ref{app B}. The kinetic mixing matrices are
\be\label{geff}
(g_{\rm eff})^{tt}_{kn,lm} =-{1\over (g_{\rm eff})^{rr}_{kn,lm}} = -\frac{2(M_{\rm eff})_{kn,\,lm}}{r-2(M_{\rm eff})_{kn,\,lm}}+O\left(\varepsilon ^3\right), 
\ee
with 
\be\label{Meff}
(M_{\rm eff})_{kn,\,lm}=M+\frac{15}{8 \pi r}\frac{{\cal I}_{kn,\,lm}  }{\Lambda_{kn,\,lm} }\varepsilon ^2
\ee
where ${\cal I}_{kn,\,lm}=\int d\Omega \; \sin^2 2\theta \; Y_k^{n*} Y_l^m$. Note that on arriving the RHS of \eq{Meff}, we have used the fact that $r\simeq 2M +{\cal O}(\varepsilon^2)$ in the near horizon region\footnote{See Appendix \ref{app B} for more details though it will not affect the main conclusion.}.

From the above we can see that the reduced near-horizon action \eq{mixingS} has high level of kinetic-mixings among spherical harmonic modes. Thus, it is impossible to interpret the kinetic mixing matrices as the reduced effective metric and determine the Hawking temperature of the scalar modes unless we can remove the mode-mixings. This is very different from the case of Schwarzschild and Kerr black holes for which the reduced action has no mode-mixing and all the modes see the same reduced metric and Hawking temperature.

As a naive try, we can truncate the kinetic mixings at some level and diagonalize the subspace of the truncated modes. For example, a two-by-two kinetic mixing matrix (e.g., $(g_{\rm eff})^{tt,rr}$ modulo an overall factor) has the following structure:
\be
 \left(
\begin{array}{cc}
 1+\alpha _0 \varepsilon +\varepsilon ^2 \mu _0\mp \frac{2 \varepsilon ^2 \zeta _0 M}{r-2 M}
   & \beta _0 \varepsilon +\varepsilon ^2 \nu _0\mp \frac{2 \varepsilon ^2 \eta _0 M}{r-2 M}
   \\
 \beta _0 \varepsilon +\varepsilon ^2 \nu _0\mp \frac{2 \varepsilon ^2 \eta _0 M}{r-2 M} &
  1+ \alpha _2 \varepsilon +\varepsilon ^2 \mu _2\mp\frac{2 \varepsilon ^2 \zeta _2 M}{r-2
   M} \\
\end{array}
\right)\nn
\ee
where $\mp$ correspond to $tt$- and $rr$-components, respectively.  Note that these two kinetic mixing matrices can be diagonalized simultaneously. After diagonalization and modulo the normalization factor, we get the reduced metrics up to ${\cal O}(\varepsilon^2)$ seen by the normal modes labelled by $\pm$ in the following form
\be
g_{\rm eff,\pm}^{tt}=-(g_{\rm eff,\pm}^{rr})^{-1}=-{2M\over r-2M}(1 \pm c_0 \varepsilon^2 {2M  \over 2M-r})
\ee
where $c_0$ is an $M$-independent constant. Using these reduced metrics we can find the Hawking temperature via surface gravity by using (1+1)d version of \eq{kappaS}  and the Hawking flux by using the method of anomaly cancellation. It is easy to see that both quantities are the same as the ones of Schwarzschild black hole of the same mass although the two modes see different metrics. 
However, this is far from the full story as we have done severe level truncation from an infinite mode-mixings. Below we remove the mode-mixings by field re-definitions.

\section{Removal of mode-mixing by field re-definitions} \label{sec 4}
We now perform a series of field re-definitions to disentangle the mode-mixings and find the reduced metric for each unmixed mode. We first introduce the shorten notation: $(A)_{lm,lm}\rightarrow (A)_{lm}$ for some quantity $A$ such as $g^{tt,rr}_{\rm eff}$ or $\Lambda^{(K)}$, then we can rewrite \eq{mixingS} into the following:
\bea
S_{n.h.}&=&4M^2 \int dt dr \sum_{l,m}  \sum_{K=0,2,4} L^{(K)}_{lm} \\
&\equiv& 4M^2 \int dt dr \sum_{l,m} L_{lm}
\eea
with
\be
L^{(K)}_{lm}= (g^{tt}_{\rm eff})_{l+K\,m}  \Lambda^{(K)}_{lm} \partial_t \phi^*_{l+K\,m} \partial_t \phi_{lm} + (t\rightarrow r).
\ee
Using the fact 
\be\label{orderEs}
\frac{(g_{\rm eff})^{tt,rr}_{l+K\,m}}{(g_{\rm eff})^{tt,rr}_{lm}}\sim 1+\varepsilon ^2,\quad 
\Lambda_{lm}^{(K)} \sim \varepsilon^{K/2},
\ee
and performing the rescaling: $\phi_{lm}\rightarrow \phi_{lm} / \sqrt{\Lambda^{(0)}_{lm}}$, then
\bea\label{middle-L}
 L_{lm}&=& (g^{tt}_{\rm eff})_{lm} \partial_t (\phi^*_{lm}+2\Gamma^*_{lm})\partial_t \phi_{lm}+ (t\rightarrow r) \nn \\
&=& (g^{tt}_{\rm eff})_{lm} \Big( |\partial_t \varphi_{lm}|^2-(\overline{\Lambda}^{(2)}_{lm})^2 |\partial_t \phi_{l+2\,m}|^2 \Big) \nn \\
&&+ (t\rightarrow r)  
\eea
with
\bea
\varphi_{lm} &\equiv&\phi_{lm}+\Gamma_{lm}, \\
\Gamma_{lm}&\equiv&\sum_{K=2,4} \overline{\Lambda}^{(K)}_{lm}\phi_{l+K\,m},\\
\overline{\Lambda}^{(K)}_{lm}&=&{1\over 2} \frac{ \Lambda^{(K)}_{lm} }{\sqrt{ \Lambda^{(0)}_{lm}  \Lambda^{(0)} _{l+K\,m}} } \sim {\cal O}(\varepsilon^{K/2}). \label{overlineL}
\eea

  If there were no second term on the RHS of \eq{middle-L}, we would have achieved the goal of removing the mode-mixing and the $\varphi_{lm}$'s will be the normal modes.  However, we can ``slide" the second term in the summation of $\sum_{lm} L_{lm}$. To proceed, we first notice that
\be
(g^{tt}_{\rm eff})_{lm} (\overline{\Lambda}^{(2)}_{lm})^2 = (g^{tt}_{\rm eff})_{l+2\,m} (\overline{\Lambda}^{(2)}_{lm})^2 + O\left(\varepsilon ^3\right)
\ee
by using \eq{orderEs} and \eq{overlineL}.  Next, we re-organize  $\sum_{lm} L_{lm}$ as follows:
\bea
\sum_{l,m} L_{lm}&=&(\sum_{l=0}^1\sum_{|m|=0}^l  + \sum_{l=2}^{\infty} \sum_{|m|=l-1}^l)    (g^{tt}_{\rm eff})_{lm} |\partial_t \varphi_{lm}|^2 \nn \\
&+& \sum_{l=2}^{\infty} \sum_{|m|=0}^{l-2}  (g^{tt}_{\rm eff})_{lm} \Big(|\partial_t \varphi_{lm}|^2 - (\overline{\Lambda}^{(2)}_{l-2\,m})^2 |\partial_t \phi_{lm}|^2 \Big)  \nn  \\
&& + (t\rightarrow r). \label{Lastpiece}
\eea 
The modes appearing in the RHS of the first line are all unmixed and the normal modes are just $\varphi_{lm}$ for $l=0,1$ with $|m|=l$ and for $l=2,\cdots,\infty$ with $|m|=l-1,l$. We thus only need to bring the RHS of the second line into the canonical form by the following field rescaling: $\phi_{lm}\rightarrow \phi_{lm}/\sqrt{1- (\overline{\Lambda}^{(2)}_{l-2\,m})^2}$ for $l=2,\cdots,\infty$ with $|m|=0,\cdots, l-2$. Besides the scaling, we should also use the fact that under this scaling, $\Gamma_{lm}^{(K)} \rightarrow \Gamma_{lm}^{(K)}+O\left(\varepsilon ^3\right)$ and $\partial_t \phi\,\Gamma_{lm}^{(K)*} \rightarrow \partial_t \phi\,\Gamma_{lm}^{(K)*}+O\left(\varepsilon ^3\right)$ for all $l$ and $m$, where $K=2,4$, then we can achieve the ``sliding" and also ensure the already canonical modes, i.e., the  $\varphi_{lm}$ in the first line of \eq{Lastpiece},  are not affected.  Without the above fact, the sliding does not work simply from the naive argument of completing the square. This indicates the subtlety of our procedure.

Note that the above procedure of field-redefinition goes almost the same for the $rr$-part.  Therefore, after all the field re-definitions and the ``slidings", we can bring the reduced action \eq{mixingS} into the following unmixed canonical form
 \be\label{canonicalS}
 S_{n.h.}=4M^2 \int dt dr \sum_{l,m} \Big( (g^{tt}_{\rm eff})_{lm} |\partial_t \varphi_{lm}|^2 + (g^{rr}_{\rm eff})_{lm} |\partial_r \varphi_{lm}|^2 \Big)
 \ee
with $(g^{tt,rr}_{\rm eff})_{lm}$ being $(g^{tt,rr}_{\rm eff})_{lm,lm}$ given in \eq{geff} and \eq{Meff}.

\section{Hawking flux \`a la anomaly cancellation}    \label{sec 5}

Based on the canonical reduced action \eq{canonicalS}, we see that each normal mode sees different metric which depend on $l,m$. Especially, the position of the horizon seen by the normal mode $\varphi_{lm}$ is at
\be
r=2 (M_{\rm eff})_{lm,lm}\equiv 2 (M_{\rm eff})_{lm}.
\ee
Despite that, we can calculate the ``Hawking temperature for each normal mode" $(T_H)_{lm}$ based on the corresponding effective reduced metric, which remarkably turns out to be mode-independent, i.e., $(T_H)_{lm}=T_H$, and is the same as the Schwarzschild's one up to ${\cal O}(\varepsilon^2)$ as
\be\label{HawkingT-reduce}
T_H=\Big({1\over 4\pi} \big|\partial_r (g^{tt}_{\rm eff,\pm})\big|\Big)\Big|_{r=2 (M_{\rm eff})_{lm}}={1\over 8 \pi M}+{\cal O}(\varepsilon^3).
\ee
This result is in consistent with the surface gravity obtained in \eq{kappaS}.  
  
  For each normal mode we consider the associated Hawking flux by the method of anomaly cancellation as for the Schwarzschild case. The only difference is now the reduced metrics are mode-dependent. For the reduced theory  \eq{canonicalS} we can write down the conservation law for both the consistent and covariant stress tensors with gravitational anomaly for each normal mode, respectively, 
\begin{align}      
\nabla_\mu T^\mu{}_{\nu,\, lm}  
&= \frac{1}{96\pi \sqrt{-(g_{\rm eff})_{lm} }}\epsilon^{\beta\delta} \partial_\delta \partial_\alpha \Gamma^\alpha_{\nu \beta,\,lm},  
\\
\nabla_\mu \widetilde{T}^\mu{}_{\nu,\, lm}   
&= - \frac{1}{96\pi \sqrt{-(g_{\rm eff})_{lm}}}\epsilon_{\mu\nu}\partial^\mu R_{lm}.
\end{align}
By adopting the method of anomaly cancellation \cite{Robinson:2005pd,Iso:2006wa,Iso:2006ut}, despite of the above anomalous Ward identity for horizon modes, we shall require the invariance of total effective action and the boundary condition by requiring the covariant stress tensor vanishes on the horizon, i.e.,   
\be
(\widetilde{T})_{\mu \nu,\,lm} \big|_{r=(r_{h\,(eff)})_{lm}}=0, 
\ee
such that there is no out-going mode from the horizon. From the above, one can deduce that the Hawking flux $a_o$ is related to the  gravitational anomaly $N^r_t$ in \eq{Nmunu} via
\be\label{hawking-flux}
 a_o=N^r_t|_{r=2 (M_{\rm eff})_{lm}}={\pi \over 12} T_H^2.
\ee
We can find that the Hawking flux is the same as the Schwarzschild's one.

\section{Discussion and Conclusion} \label{sec 6}
By exploiting the anomaly cancellation method, we  in this paper have shown the Hawking flux for the non-rotating black hole with quadratic supertranslation hair is the same as the Schwarzschild's one 
even though the non-trivial mode-mixings arise in the near-horizon reduced field theories due to the non-spherical shape of the horizon. For concreteness we have chosen a particular soft hair $C$ 
function (i.e., $Y_2^0(\theta,\phi)$) to proceed our calculations, however  our conclusion should hold for the more generic quadratic supertranslation hairs\footnote{We indeed have tried the quadratic 
soft hair with the simpler function $C=\epsilon M Y^0_1(\theta,\phi)$ and find that the Hawking flux is the same as the ones for Schwarzschild black hole, though we do not give the details here.}. This 
is easy to be appreciated by the following facts: (1) as long as the soft hair $C$ function is $\phi$-independent but still $\theta$-dependent, the near-horizon metric should still takes the form of 
\eq{gtt-nh}-\eq{gphiphi-nh} but with different sub-leading coefficients of $\epsilon$ expansions. (2) These changes will only affect the mode coefficients ${\cal I}^{A_i,B_i,C_i,D_i}_{lm}$ given in 
Appendix \ref{app B}, which will then affect the mode-dependent part of the reduced metric \eq{geff}. (3) However, since the Hawking flux 
\eq{hawking-flux} does not depend on the details of the mode-dependent part of the reduced metric, we will expect to get the same result. (4) 
Thus, our result should be universal although it needs further study if the $C$ function is also $\phi$-dependent.

Lastly, our results obtained in this study are non-trivial and against expectations. 
The reason for ``non-trivial'' is that our result can be obtained through the highly complicated calculation as in section \ref{sec 4},
field re-definitions and disentanglement of mode-mixings.  
The reason for ``against expectations'' is that 
supertranslation and superrotation map one spacetime solution of Einstein equation to another solution, and they differ by the soft hairs which are the 
associated charges of supertranslation and superrotation. The values of the soft hairs are determined from the process of the black hole formation, thus the soft hairs  
are considered to correspond to the microscopic degrees of freedom encoded in the spacetime of a black hole  \cite{Compere:2016hzt, Hawking:2016sgy}. Naively one may 
expect that the Hawking temperature and Hawking flux will be modified by the addition of the degrees of freedom associated with soft hairs.

Instead, our results show that the Hawking temperature and flux are invariant under supertranslations, which means the Hawking temperature and flux are independent 
of the supertranslation hairs, at least up to the quadratic order, which may be related to the fact that the black hole mass (measured as the Bondi mass) is invariant under 
supertranslations. This may also be the case for the superrotations. Therefore, our results imply that the nonlinear soft hairs can be thought as the degrees of freedom 
associated with the black hole microstates, which may change the shape of the horizon but not the macroscopic thermodynamical quantities such as the Hawking temperature and flux.

\subsection*{Acknowledgements}
S.T. thank Wen-Yu Wen for his discussion very much. 
FLL is supported by Taiwan Ministry of Science and Technology (MoST) through Grant No.~106-2112-M-003-004-MY3, and he also thank NCTS for partial financial support. 


\allowdisplaybreaks

\appendix

\section{Metric of supertranslated black hole}\label{app A}

In this Appendix we give some details about the metric we adopt for our study.

By adopting the soft hair function $C = \varepsilon M Y_2^0 (\theta,\phi)= \varepsilon M \sqrt{\frac{5}{16\pi}}\left(3\cos^2\theta-1\right)$ (see also \eq{Cfunction} in the main text),  the metric with quadratic supertranslation hair we will study take the following form in the isotropic coordinate
\be\label{startmetric}
ds^2=-\frac{\left( 1 - \frac{M}{2\rho_s} \right)^2}{\left( 1 + \frac{M}{2\rho_s} \right)^2}dt^2 +\left( 1 + \frac{M}{2\rho_s} \right)^4(d\rho^2+ \rho^2 d\tilde{\Omega}_2)  
\ee
where the radial coordinate $\rho$ is related to $\rho_s$ by
\begin{eqnarray} \label{ctki21}
\rho_s^2 
= 
\frac
{45 \varepsilon ^2 M^2 \sin ^2(2 \theta )}
{16 \pi }
+\left(
\rho 
-\frac{1}{8} \sqrt{\frac{5}{\pi }} 
\varepsilon  M 
\left(
3 \cos 2 \theta +1
\right)
\right)^2, \nn
\end{eqnarray} 
and the metric of the deformed 2-sphere is given by
\be
d\tilde{\Omega}_2= (1+\epsilon_{\theta\theta})d\theta^2  + \sin^2\theta (1+\epsilon_{\phi\phi}) d\phi^2
\ee
with
\begin{eqnarray}
\label{ctkitt}  
\epsilon_{\theta\theta}
&=& \frac{1}{4 \rho} \sqrt{\frac{5}{\pi }} \varepsilon  M (9 \cos (2 \theta )-1)
+\frac{5 \varepsilon ^2 M^2 (1-9 \cos (2 \theta ))^2}{64 \pi \rho^2}  \nn \\
&& +O\left(\varepsilon ^3\right), \nn
\\*
\label{ctkipp} 
\epsilon_{\phi\phi}
&=&
{1\over 4\rho}  \sqrt{5 \over \pi } \varepsilon  M  (3 \cos (2 \theta )+5)
+{5 \over 64 \pi} \varepsilon ^2 M^2 (3 \cos (2 \theta )+5)^2
\nonumber\\*
&&+O\left(\varepsilon ^3\right). \nn
\end{eqnarray} 

   Next, we like to rewrite the metric \eq{startmetric} into the form in the Schwarzschild coordinate by requiring
 \bea
& \left( 1-\frac{2 \mu(\rho) }{r} \right) = \frac{\left( 1 - \frac{M}{2\rho_s} \right)^2}{\left( 1 + \frac{M}{2\rho_s} \right)^2},
\\
& \frac{1}{1-\frac{2 \mu(\rho) }{r}} \left( \frac{dr}{d\rho} \right)^2 = \left( 1 + \frac{M}{2\rho_s} \right)^4
 \eea
so that the $tt$ and $rr$-part of the resultant metric will take the following form
\be
-\left(1-\frac{2\mu(\rho)}{r}\right)dt^2+\frac{1}{1-\frac{2\mu(\rho)}{r}}dr^2+\cdots. \nn
\ee
Solving the above conditions for the Schwarzschild radial coordinate $r$ and $\mu(\rho)$, and use the solutions to perform the coordinate transformations to Schwarzschild coordinate. After lengthy calculations, we arrive the following metric in Schwarzschild coordinate
\be
ds^2=g_{\mu\nu} dx^{\mu}dx^{\nu}
\ee
with nonzero components given as follows:
\begin{align}
g_{tt} =& 
-\left(1-\frac{2 M}{r}\right) \nn
\\
& +
\frac{
15 \varepsilon ^2 \sin ^2 2 \theta 
\left(M^5-3 M^3 \left(\sqrt{r (r-2 M)}- M + r\right)^2\right)
}{
4 \pi  r \left(r + \sqrt{r (r-2 M)}\right)^2 
\left(\sqrt{r (r-2M)}- M+r\right)^2
} \nn \\
& +O\left(\varepsilon ^3\right), \nn
\end{align}

\begin{align}
g_{rr} 
=&
\frac{1}{1-\frac{2 M}{r}}  
+
\frac{15 \varepsilon ^2 \sin ^2 2 \theta  }
{4 \pi  M (r-2M)^2}
\bigg\{
M^3+3 M^2 r-6 M r^2
\nonumber\\*
&
- 2 \left(\sqrt{r^5 (r-2 M)}- r^3\right)+ 4 M r \sqrt{r (r-2 M)}
\bigg\}
+O\left(\varepsilon ^3\right), \nn
\end{align}

\begin{align}
g_{\theta\theta} =& 
r^2
+\frac{3 \sqrt{\frac{5}{\pi }} \varepsilon  M \cos 2 \theta \left(\sqrt{r (r-2M)}+r\right)^4}{2 \left(\sqrt{r (r-2M)}-M+r\right)^3}
+{N_{\theta} \over D_{\theta}} +O\left(\varepsilon ^3\right), \nn
\end{align}
with 
\begin{align}
D_{\theta}=& 8 \pi  r (r-2M)^2  \Big[ M^4-32 M r^3-24 M \sqrt{r^5 (r-2M)} \nn 
\\
& +20 M^2 r \left(2 r+\sqrt{r (r-2M)}\right) -4 M^3 \left(4 r+\sqrt{r (r-2M)}\right) \nn
\\
&+8 \left(r^4+\sqrt{r^7 (r-2M)}\right)  \Big]  \nn
\end{align}
and
\begin{align}
N_{\theta} =&15 \varepsilon ^2 M^2 \Big[ -4 M^5 r^2+8 M^4 \left(2 \sqrt{r^5 (r-2 M)}+13 r^3\right) \nn \\
& -5 M^3 \left(32 \sqrt{r^7 (r-2M)}+73 r^4\right) \nn \\
&  +M^2 \left(268 \sqrt{r^9 (r-2M)}+409r^5\right) \nn \\
& -6 M \Big(26 \sqrt{r^{11} (r-2M)} +31r^6\Big) \nn\\
& +30 \Big(\sqrt{r^{13} (r-2M)}+r^7 \Big) \nn\\
& + \cos 4 \theta \Big\{ 4 M^5 r^2-8 M^4 \left(2 \sqrt{r^5 (r-2 M)}+r^3\right) \nn \\
& +M^3 \left(-32 \sqrt{r^7 (r-2M)}-115 r^4\right) \nn \\
& +M^2 \left(116 \sqrt{r^9 (r-2M)}+191r^5\right)-102M r^6 \nn \\
& -84 M \sqrt{r^{11} (r-2M)}+18 \left(\sqrt{r^{13} (r-2M)}+r^7\right) \Big\} \Big], \nn
\end{align}
and 
\begin{align}
g_{r\theta} 
=&
-\frac{3 \sqrt{\frac{5}{\pi }} \varepsilon  M \sin 2 \theta \left(r+\sqrt{r (r-2M)}\right)^4}{8 \sqrt{r (r-2M)} \left(\sqrt{r (r-2M)} - M + r\right)^3}\nn \\
&
+ \frac{15 \varepsilon ^2 M^4 r \sin 4 \theta}{4 \pi  (r-2M) \left(\sqrt{r (r-2M)} - M + r\right)^3}
+O\left(\varepsilon ^3\right), \nn
\end{align}

\begin{align}
g_{\phi\phi} =& 
r^2 \sin ^2 \theta
+\frac{3 \sqrt{\frac{5}{\pi }} \varepsilon  M \sin ^2 2 \theta \left(\sqrt{r (r-2M)}+r\right)^4}{8 \left(\sqrt{r (r-2M)}-M+r\right)^3}\nonumber  \nn\\
& +\frac{15 \varepsilon ^2 M^2 \sin^2 2\theta }{8 \pi  r^2 (r-2M)^2 \left(\sqrt{r (r-2M)}-M+r\right)^4}  H_{\phi} \nn \\&+O\left(\varepsilon ^3\right), \nn
\end{align}
with
\begin{align}
H_{\phi}=& -4 M^5 r^3+8 M^4 \left(2 \sqrt{r^7 (r-2M)}+7 r^4\right)\nonumber\\*
&-M^3 \left(64 \sqrt{r^9 (r-2M)}+125 r^5\right) \nn \\
&+M^2 \left(76\sqrt{r^{11} (r-2M)}+109 r^6\right)\nn \\
&-6 M \left(6 \sqrt{r^{13} (r-2M)}+7 r^7\right) \nn \\
&+6 \left(\sqrt{r^{15} (r-2M)}+r^8\right)\nn \\
&+\Big\{4 M^5 r^3-16 M^4 \left(\sqrt{r^7 (r-2M)}+2 r^4\right)\nn \\
&+M^3 \left(16 \sqrt{r^9 (r-2 M)}+5 r^5\right)\nonumber\\*
&+M^2 \left(20\sqrt{r^{11} (r-2M)}+41 r^6\right)\nn \\
&-6 M \left(4 \sqrt{r^{13} (r-2M)}+5 r^7\right) \nn \\
&+6 \left(\sqrt{r^{15} (r-2M)}+r^8\right)\Big\} \cos 2 \theta.  \nn 
\end{align}

From the above, we can solve the position of the horizon $r_h$ from the condition $g_{tt}(r_h)=0$, and the result is
\begin{align}
r_h=2 M-\frac{15 \varepsilon ^2 M \sin ^2(2 \theta )}{8 \pi }+O\left(\varepsilon ^3\right).
\end{align}
We see that $r_h$ receive an angular-dependent correction to $r=2M$ at $O(\varepsilon^2)$. This angular correction of the horizon position results in the nontrivial mode-mixing when performing the dimensional reduction. 

Since we are only interested in the near-horizon region when considering the Hawking flux by the anomaly cancellation method, we will further take the near horizon limit of the above metric, namely, taking $r\simeq r_h$ and keep up to the leading order of ${\cal O}(r-r_h)$. The resultant near-horizon metric is 
\be 
ds^2=g^{(n.h.)}_{\mu\nu} dx^{\mu}dx^{\nu} \nn
\ee
with the nonzero components up to $O(\varepsilon^2)$ as follows:
\bea 
g^{(n.h.)}_{tt}&=&-(g^{(n.h.)}_{rr})^{-1}=-{r-r_h \over 2M} \left(1-{45 \varepsilon^2 \sin^2 2\theta \over 8\pi}\right), \nn
\\
g^{(n.h.)}_{\theta\theta}&=& r^2 \Big(1+ 6\sqrt{5\over \pi} \varepsilon \cos 2\theta +{5\varepsilon^2 \over 16\pi} (75+69 \cos 4\theta)\Big), \nn
\\
g^{(n.h.)}_{r\theta}&=&-3\varepsilon \sqrt{10 M^3 \over \pi (r-r_h)} \sin 2\theta + {15 \varepsilon^2 M^2 \sin 4\theta \over 2\pi (r-r_h)}, \nn 
\\
g^{(n.h.)}_{\phi\phi}&=&r^2 \sin ^2 \theta \Big(1+ 3 \varepsilon \cos^2 \theta \Big\{ 2\sqrt{5\over \pi} + {5\over 4 \pi} \varepsilon \nn \\
&&- {35 \over 8\pi}  (r-r_h) \varepsilon + {55\over 4\pi} \varepsilon \cos 2\theta \Big\} \Big). \nn
\eea

\section{Reduced action}\label{app B}

In this Appendix we give some more details about the dimensionally reduced action. 

The more detailed form of the reduced action of \eq{mixingS} in the main text is 
\begin{eqnarray}
&& 
\sum_{kn} \sum_{lm} \int d^2x\,  \Bigg\{
\phi_{kn}^* \Big( 
-\frac{2M}{r-2M} \Lambda_{lm,\,kn}\nn \\
&& -\frac{15M^2\varepsilon^2}{4\pi(r-2M)^2} {\cal I}^{C}_{kn,\,lm}
\Big) \partial_t\partial_t\phi_{lm}
\nonumber\\*
&&
+\,
\phi_{kn}^* \partial_r
\left(
\frac{r-2M}{2M} \Lambda_{lm,\,kn} -\frac{15M^2\varepsilon^2}{16\pi} {\cal I}^{C}_{kn,\,lm}
\right)  
\partial_r\phi_{lm}
\Bigg\}
\nonumber\\*
\label{cagttpteg}
&\equiv& 
\sum_{kn} \sum_{lm}
\int d^2x\, \Lambda_{kn,\,lm} 
\Big(g_{\rm eff})^{tt}_{kn,\,lm}  \partial_t \phi^*_{kn} \partial_t \phi_{lm}\nn \\
&& +(g_{\rm eff})^{rr}_{kn,\,lm}  \partial_r \phi^*_{kn} \partial_r \phi_{lm} \Big) \nn
\end{eqnarray} 
with 
\be\label{geff-app}
(g_{\rm eff})^{tt}_{kn,lm} =\left(-(g_{\rm eff})^{rr}_{kn,lm} \right) ^{-1}\equiv
-\frac{2M}{r-2 M-\frac{{\cal I}^{C}_{kn,\,lm}}{\Lambda_{kn,\,lm}}\frac{15 M \varepsilon ^2 }{8 \pi r}}  
\ee
and 
\begin{align}
\Lambda_{kn,\,lm} 
=& \, 
4M^2 \Bigg\{ \Bigg( 
1+\frac{3}{2}\sqrt{\frac{5}{\pi}}\left(1+3{\cal I}^{A_0}_{lm}\right)\varepsilon
\nn \\
&+\frac{45}{2\pi}\left(-{\cal I}^{B_0}_{lm}+3{\cal I}^{D_0}_{lm}\right)\varepsilon^2 
\Bigg)
\delta_{kl}\delta_{mn}
\nonumber \\*[0.5mm]
&
+\frac{9}{2} 
\left( 
\sqrt{\frac{5}{\pi}} {\cal I}^{A_2}_{lm} \varepsilon
+\frac{5}{\pi}\left( -{\cal I}^{B_2}_{lm}+3{\cal I}^{D_2}_{lm} \right)\varepsilon^2 
\right) 
\delta_{k-2,l}\delta_{mn} \nn \\
&+\frac{135}{2\pi} {\cal I}^{D_4}_{lm} \varepsilon^2 
\delta_{k-4,l}\delta_{mn}  \Bigg\}
\nonumber\\*[0.5mm]
\label{ldcep} 
\equiv& \, 
\Lambda_{lm}^{(0)} \delta_{kl}\delta_{mn} + \Lambda_{lm}^{(2)}  \delta_{k-2,l}\delta_{mn} + \Lambda_{lm}^{(4)}  \delta_{k-4,l}\delta_{mn}. \nn
\end{align}
where 
\begin{align}
& 
\int d\Omega \cos 2\theta \,(Y_k^n)^* Y_l^m 
\nonumber\\*
=& 
-\frac{4 m^2-1} {4 l^2+4 l-3}\delta _{kl}\delta _{nm} \nn \\
&+\frac{2 (-1)^{2 m} 
\sqrt{\frac{((l+1)^2-m^2) ((l+2)^2-m^2)}{(2 l+1) (2 l+5)}} }{2 l+3}\delta _{k-2,l}\delta _{nm} 
\nonumber\\*[2.5mm]
\equiv& \, {\cal I}^{A_0}_{lm}\,\delta _{kl} \delta _{nm}+{\cal I}^{A_2}_{lm}\,\delta _{k-2,l} \delta _{nm}
\equiv{\cal I}^{A}_{lm,\,kn}, \nn
\end{align}
\begin{align}
&
\int d\Omega \cos^2 \theta \,(Y_k^n)^* Y_l^m 
\nonumber\\* 
=& 
\frac{2 l^2+2 l-2 m^2-1}{4 (l+1)^2-4 (l+1)-3}\delta _{kl}\delta _{nm} \nn \\
&+\frac{(-1)^{2 m} \sqrt{\frac{((l+1)^2-m^2) ((l+2)^2-m^2)}{(2 l+1) (2 l+5)}} }{2 l+3}\delta _{k-2,l}\delta _{nm} 
\nonumber\\*[2.5mm]
\equiv& \, {\cal I}^{B_0}_{lm}\,\delta _{kl} \delta _{nm}+{\cal I}^{B_2}_{lm} \,\delta _{k-2,l}\delta _{nm}
\equiv{\cal I}^{B}_{lm,\,kn},  \nn
\end{align}

\begin{align}
&
\int d\Omega \sin^2 (2\theta) \, (Y_k^n)^* Y_l^m 
& \nonumber\\*
=&8 (-1)^{2 m}  \delta _{kl} \delta _{nm} \times \nn \\
& \frac{\left(l (l+1)\left(l^2+l-5\right)+2 l (l+1) m^2-3 m^4+3\right)}{(2 l-3) (2 l-1) (2 l+3) (2 l+5)} 
\nonumber\\*
&+4 (-1)^{2 m-2 l} \delta _{k-2,l}\delta _{nm} \sqrt{\frac{((l+1)^2-m^2) ((l+2)^2-m^2)}{4 l^2+12 l+5}} 
\nonumber\\*
& \times \frac{(-1)^{2 l} (2 l-1) (2 l+7)-4 (-1)^{2 m} \left(l (l+3)-7 m^2\right) }{7 (2 l-1) (2 l+3) (2 l+7)} 
\nonumber\\*
&-4 (-1)^{2m} \sqrt{\frac{((l+1)^2-m^2) ((l+2)^2-m^2)}{(2 l+1) (2 l+3)^2 (2 l+9)}} \nn \\
&\times \sqrt{\frac{((l+3)^2-m^2) ((l+4)^2-m^2) }{(2 l+5)^2 (2 l+7)^2}} \delta _{k-4,l}\delta _{nm} 
\nonumber\\*[2.5mm]
\equiv&  \,{\cal I}^{C_0}_{lm}\,\delta _{kl} \delta _{nm}+{\cal I}^{C_2}_{lm} \,\delta _{nm} \delta _{k-2,l}+{\cal I}^{C_4}_{lm} \,\delta _{nm} \delta _{k-4,l}
\equiv{\cal I}^{C}_{lm,\,kn},  \nn
\end{align}

\begin{align}
&
\int d\Omega \cos 2\theta \cos^2 \theta\, (Y_k^n)^* Y_l^m 
& \nonumber\\*
=& \frac{(-1)^{2 m}  \delta _{kl} \delta _{nm} }{(2 l-3) (2 l-1) (2 l+3) (2 l+5)} \Big(12m^4+30 m^2+3 \nn \\
& +2 \left(-8 l (l+1) m^2+l(l+1) (2 l (l+1)-7)\right) \Big) \nn\\
&+(-1)^{2 m-2 l} \delta _{k-2,l} \delta _{nm}\sqrt{\frac{((l+1)^2-m^2) ((l+2)^2-m^2)}{4 l^2+12 l+5}} 
\nonumber\\*
&\times\frac{ 8 (-1)^{2 m} \left(l (l+3)-7 m^2\right)+5 (-1)^{2 l} (2 l-1) (2 l+7)}{7 (2 l-1) (2 l+3) (2 l+7)} 
\nonumber\\*
& +2 (-1)^{2m} \delta _{k-4,l}\delta _{nm} \sqrt{\frac{((l+1)^2-m^2) ((l+2)^2-m^2)}{(2 l+1) (2 l+3)^2 (2 l+9)}} \nn \\
&\times \sqrt{\frac{((l+3)^2-m^2) ((l+4)^2-m^2) }{(2 l+5)^2 (2 l+7)^2}} 
\nonumber\\*[2.5mm]
\equiv&  \, {\cal I}^{D_0}_{lm} \, \delta _{kl} \delta _{nm}+{\cal I}^{D_2}_{lm}\, \delta _{nm} \delta _{k-2,l}+{\cal I}^{D_4}_{lm} \, \delta _{nm} \delta _{k-4,l}
\equiv{\cal I}^{D}_{lm,\,kn}. \nn
\end{align}

Note that we have used the contraction rule of spherical harmonics to relate the product of two spherical harmonics to the Clebsch-Gordan coefficients, i.e.,
\begin{align}  
\vspace{20mm}
& \int d\Omega~ (Y_{l_1}^{m_1})^* (Y_{l_2}^{m_2})^* Y_L^M = \sqrt{\frac{(2l_1+1)(2l_2+1)}{4\pi (2L+1)}} 
\nonumber  
\\
& \times\langle l_1 0~l_2 0| L  0\rangle \langle l_1 m_1~l_2 m_2| LM \rangle,
\end{align}
and then evaluate the above integrals explicitly.  However, as shown in the main text, the final answer for the Hawking flux will not depend on the details of ${\cal I}^{A_i,B_i,C_i,D_i}_{lm}$ so that our result should be quite universal for the general quadratic supertranslation hair. 

  We comment that we can further simplify the reduced metric given in \eq{geff-app}  by approximating $r=2M+{\cal O}(\varepsilon^2)$, and then arrive the form given in \eq{geff} and \eq{Meff} of the main text.  However, the final result for the Hawking flux will be the same for either using this approximation or not.


\begin{thebibliography}{99}%

\bibitem{Mathur:2009hf} 
  S.~D.~Mathur,
  ``The Information paradox: A Pedagogical introduction,''
  Class.\ Quant.\ Grav.\  {\bf 26}, 224001 (2009)
  [arXiv:0909.1038 [hep-th]].
  
\bibitem{Hawking:1974rv} 
  S.~W.~Hawking,
  ``Black hole explosions,''
  Nature {\bf 248}, 30 (1974).

\bibitem{Hawking:1974sw} 
  S.~W.~Hawking,
  ``Particle Creation by Black Holes,''
  Commun.\ Math.\ Phys.\  {\bf 43}, 199 (1975)
  Erratum: [Commun.\ Math.\ Phys.\  {\bf 46}, 206 (1976)].
 
\bibitem{Bekenstein:1973ur} 
  J.~D.~Bekenstein,
  ``Black holes and entropy,''
  Phys.\ Rev.\ D {\bf 7}, 2333 (1973).
  
\bibitem{Hawking:2016msc} 
  S.~W.~Hawking, M.~J.~Perry and A.~Strominger,
  ``Soft Hair on Black Holes,''
  Phys.\ Rev.\ Lett.\  {\bf 116}, no. 23, 231301 (2016)
  [arXiv:1601.00921 [hep-th]].

\bibitem{Strominger:2014pwa}
A.~Strominger and A.~Zhiboedov,
``Gravitational Memory, BMS Supertranslations and Soft Theorems,''
JHEP \textbf{01}, 086 (2016)
doi:10.1007/JHEP01(2016)086
[arXiv:1411.5745 [hep-th]].
 
\bibitem{Strominger:2017zoo}
A.~Strominger,
``Lectures on the Infrared Structure of Gravity and Gauge Theory,''
[arXiv:1703.05448 [hep-th]].

\bibitem{Bondi:1962px} 
  H.~Bondi, M.~G.~J.~van der Burg and A.~W.~K.~Metzner,
  ``Gravitational waves in general relativity. 7. Waves from axisymmetric isolated systems,''
  Proc.\ Roy.\ Soc.\ Lond.\ A {\bf 269}, 21 (1962).

\bibitem{Sachs:1962wk} 
  R.~K.~Sachs,
  ``Gravitational waves in general relativity. 8. Waves in asymptotically flat space-times,''
  Proc.\ Roy.\ Soc.\ Lond.\ A {\bf 270}, 103 (1962).
  
\bibitem{Sachs:1962zza} 
  R.~Sachs,
  ``Asymptotic symmetries in gravitational theory,''
  Phys.\ Rev.\  {\bf 128}, 2851 (1962).
  
  
\bibitem{Braginsky:1986ia}
V.~Braginsky and L.~Grishchuk,
``Kinematic Resonance and Memory Effect in Free Mass Gravitational Antennas,''
Sov.\ Phys.\ JETP \textbf{62}, 427-430 (1985).
  
\bibitem{Braginsky&Thorne} 
V. B. Braginsky and K. S. Thorne, ``Gravitational-wave bursts with memory and experimental prospects", Nature 327 (1987) 123.  
  
\bibitem{Blanchet:1992br}
L.~Blanchet and T.~Damour,
``Hereditary effects in gravitational radiation,''
Phys.\ Rev.\ D \textbf{46}, 4304-4319 (1992). 
  
 
\bibitem{Mirbabayi:2016axw} 
  M.~Mirbabayi and M.~Porrati,
  ``Dressed Hard States and Black Hole Soft Hair,''
  Phys.\ Rev.\ Lett.\  {\bf 117}, no. 21, 211301 (2016)
  [arXiv:1607.03120 [hep-th]].


\bibitem{Bousso:2017dny} 
  R.~Bousso and M.~Porrati,
  ``Soft Hair as a Soft Wig,''
  Class.\ Quant.\ Grav.\  {\bf 34}, no. 20, 204001 (2017)
  [arXiv:1706.00436 [hep-th]].

\bibitem{Javadinazhed:2018mle}
R.~Javadinezhad, U.~Kol and M.~Porrati
``Comments on Lorentz Transformations, Dressed Asymptotic States and Hawking Radiation,''
JHEP \textbf{01}, 089 (2019)
[arXiv:1808.02987 [hep-th]].
  


   
  

\bibitem{Hawking:2016sgy} 
  S.~W.~Hawking, M.~J.~Perry and A.~Strominger,
  ``Superrotation Charge and Supertranslation Hair on Black Holes,''
  JHEP {\bf 1705}, 161 (2017)
  [arXiv:1611.09175 [hep-th]].
  
  
\bibitem{Compere:2016hzt} 
  G.~Compere and J.~Long,
  ``Classical static final state of collapse with supertranslation memory,''
  Class.\ Quant.\ Grav.\  {\bf 33}, no. 19, 195001 (2016)
  [arXiv:1602.05197 [gr-qc]].




\bibitem{Parikh:1999mf} 
  M.~K.~Parikh and F.~Wilczek,
  ``Hawking radiation as tunneling,''
  Phys.\ Rev.\ Lett.\  {\bf 85}, 5042 (2000)
  [hep-th/9907001].

 
\bibitem{Christensen:1977jc}   
  S.~M.~Christensen and S.~A.~Fulling,
  ``Trace Anomalies and the Hawking Effect,''
  Phys.\ Rev.\ D {\bf 15}, 2088 (1977).


\bibitem{Robinson:2005pd} 
  S.~P.~Robinson and F.~Wilczek,
  ``A Relationship between Hawking radiation and gravitational anomalies,''
  Phys.\ Rev.\ Lett.\  {\bf 95}, 011303 (2005)
  [gr-qc/0502074].

\bibitem{Iso:2006wa}
S.~Iso, H.~Umetsu and F.~Wilczek,
``Hawking radiation from charged black holes via gauge and gravitational anomalies,''
Phys. Rev. Lett. \textbf{96}, 151302 (2006)
[arXiv:hep-th/0602146 [hep-th]].


\cite{Iso:2006ut}
\bibitem{Iso:2006ut}
S.~Iso, H.~Umetsu and F.~Wilczek,
``Anomalies, Hawking radiations and regularity in rotating black holes,''
Phys. Rev. D \textbf{74}, 044017 (2006)
[arXiv:hep-th/0606018 [hep-th]].


  
 








  
\bibitem{AlvarezGaume:1983ig}
L.~Alvarez-Gaume and E.~Witten,
``Gravitational Anomalies,''
Nucl.\ Phys.\ B \textbf{234}, 269 (1984).

\bibitem{Bertlmann:2000da} 
  R.~A.~Bertlmann and E.~Kohlprath,
  ``Two-dimensional gravitational anomalies, Schwinger terms and dispersion relations,''
  Annals Phys.\  {\bf 288}, 137 (2001)
  [hep-th/0011067].
  
  
\bibitem{Chu:2018tzu} 
  C.~S.~Chu and Y.~Koyama,
  ``Soft Hair of Dynamical Black Hole and Hawking Radiation,''
  JHEP {\bf 1804}, 056 (2018)
  [arXiv:1801.03658 [hep-th]].
  

\bibitem{Compere:2019rof} 
  G.~Compre, J.~Long and M.~Riegler,
  ``Invariance of Unruh and Hawking radiation under matter-induced supertranslations,''
  JHEP {\bf 1905}, 053 (2019),
  [arXiv:1903.01812 [hep-th]].
  
\bibitem{Takeuchi:2021ibg}
S.~Takeuchi,
``Hawking flux of 4D Schwarzschild blackhole with supertransition correction to second-order,''
SciPost Phys. Proc. \textbf{4}, 010 (2021)
[arXiv:2104.05483 [hep-th]].

 
\bibitem{Berti:2005ys} 
  E.~Berti, V.~Cardoso and C.~M.~Will,
  ``On gravitational-wave spectroscopy of massive black holes with the space interferometer LISA,''
  Phys.\ Rev.\ D {\bf 73}, 064030 (2006)
  [gr-qc/0512160].



\bibitem{Townsend:1997ku} 
  P.~K.~Townsend,
  ``Black holes: Lecture notes,''
  gr-qc/9707012.
  
  
  

  
 

 \end{thebibliography}
 \end{document}